\documentclass[twocolumn]{bmcart}
\RequirePackage{hyperref}
\usepackage[utf8]{inputenc} 
\usepackage{graphicx}

\startlocaldefs
\endlocaldefs

\begin{document}

\begin{frontmatter}

\begin{fmbox}
\dochead{Research}


\title{Levels and trends in the sex ratio at birth in seven provinces of Nepal between 1980 and 2016 with probabilistic projections to 2050: a Bayesian modeling approach}


\author[
   addressref={aff1},                   
   corref={aff1},                       
   email={fengqing.chao@kaust.edu.sa}   
]{\inits{FC}\fnm{Fengqing} \snm{Chao}}
\author[
   addressref={aff2,aff3},
   noteref={n1},                        
   email={kcsamir@gmail.com}
]{\inits{SKC}\fnm{Samir} \snm{K.C.}}
\author[
   addressref={aff1},                   
   email={hernando.ombao@kaust.edu.sa}   
]{\inits{HO}\fnm{Hernando} \snm{Ombao}}


\address[id=aff1]{
  \orgname{Statistics Program, Computer, Electrical and Mathematical Sciences and Engineering Division, King Abdullah University of Science and Technology}, 
  \street{4700 KAUST},                     %
  \postcode{23955-6900}                                
  \city{Thuwal},                              
  \cny{Saudi Arabia}                                    
}
\address[id=aff2]{%
  \orgname{Asian Demographic Research Institute, Shanghai University},
  \street{Shangda Road},
  \postcode{200444}
  \city{Shanghai},
  \cny{China}
}
\address[id=aff3]{%
  \orgname{Wittgenstein Centre for Demography and Global Human Capital (IIASA, VID/OeAW, UV), International Institute for Applied Systems Analysis},
  \street{Schlossplatz 1},
  \postcode{2361}
  \city{Laxenburg},
  \cny{Austria}
}


\begin{artnotes}
\note[id=n1]{Second corresponding author: \href{mailto:kcsamir@gmail.com}{kcsamir@gmail.com}} 
\end{artnotes}



\begin{abstractbox}

\begin{abstract} 
\parttitle{Background} 
The sex ratio at birth (SRB; ratio of male to female births) in Nepal has been reported without imbalance on the national level. However, the national SRB could mask the disparity within the country. Given the demographic and cultural heterogeneities in Nepal, it is crucial to model Nepal SRB on the subnational level. Prior studies on subnational SRB in Nepal are mostly based on reporting observed values from surveys and census, and no study has provided probabilistic projections. We aim to estimate and project SRB for the seven provinces of Nepal from 1980 to 2050 using a Bayesian modeling approach.

\parttitle{Methods}
We compiled an extensive database on provincial SRB of Nepal, consisting 2001, 2006, 2011, and 2016 Nepal Demographic and Health Surveys and 2011 Census. We adopted a Bayesian hierarchical time series model to estimate and project the provincial SRB, with a focus on modelling the potential SRB imbalance.

\parttitle{Results}
In 2016, the highest SRB is estimated in Province 5 at 1.102 with a 95\% credible interval (1.044, 1.127) and the lowest SRB is in Province 2 at 1.053 (1.035, 1.109). During 1980--2016, the provincial SRB was around the same level as the national SRB baseline of 1.049. The SRB imbalance probabilities in all provinces are generally low and vary from 16\% in Province 2 to 81\% in Province 5. SRB imbalances are estimated to have begun at the earliest in 2001 in Province 5 with a 95\% credible interval (1992, 2022) and the latest in 2017 (1998, 2040) in Province 2. We project SRB in all provinces to begin converging back to the national baseline in the mid-2030s. By 2050, the SRBs in all provinces are projected to be around the SRB baseline level.

\parttitle{Conclusion}
Our findings imply that the majority of provinces in Nepal have low risks of SRB imbalance for the period 1980--2016. However, we identify a few provinces with higher probabilities of having SRB inflation. The projected SRB is an important illustration of potential future prenatal sex discrimination and shows the need to monitor SRB in provinces with higher possibilities of SRB imbalance.
\end{abstract}




\end{abstractbox}

\end{fmbox}

\end{frontmatter}



\section*{Introduction}
Under natural circumstances, the sex ratio at birth (SRB), the ratio of male to female live births, fluctuates within a narrow range, from 1.03 (i.e. 103 male births per 100 female births) to 1.07.\cite{763172:18187634,763172:18187905,763172:18187947,763172:18187948,763172:18187990,763172:18188032,763172:18188033,763172:18188043,763172:18188044,763172:18188165,763172:18188166,763172:18188167,763172:18188168,763172:18188210} However, since the 1970s, skewed and masculinized SRB has been reported in several countries worldwide, but mainly concentrated in Asia and Eastern Europe.\cite{763172:18188419,763172:18188437,763172:18188438,763172:18188439,763172:18188440,763172:18188441,763172:18188483,763172:18188486,763172:18188487,763172:18188488,763172:18188489,763172:18188490,763172:18188573,763172:18188574,763172:18188657,763172:18188658,763172:18188659,763172:18188660,763172:18188661,763172:18188703,763172:18188745,763172:18188746} SRB imbalance is a direct result of sex-selective abortion; in short, ``sex selection''. The action of sex selection in a certain population is largely driven by three factors.\cite{763172:18188490,763172:18188574} First, the population has a strong and sustained preference for sons. Second, technologies offering prenatal sex determination and abortion are accessible and affordable. Third, the number of children a woman gives birth to is decreasing as a result of fertility decline.\cite{763172:19211004} This fertility transition leads to a ``fertility squeeze'', in which sex selection is used to obtain a desired small family and ensures male offspring.

In Nepal, the three main factors leading to sex selection have been documented. First, the predominant religion in Nepal is Hinduism, and being a patriarchal society, it attributes higher value to sons than daughters. This son preference in Nepal is reflected in various aspects: a stronger desire to have sons rather than daughters;\cite{763172:18187947} among women who decide to stop or have ceased childbearing, a higher proportion of the last born children are males rather than females;\cite{763172:18188975,763172:18187947} a skewed sex ratio for a higher order of births given previous female births;\cite{763172:18188972,763172:18189027} and better access to education for boys compared to their sisters.\cite{763172:18188973} Previous studies have also found higher-than-expected mortality rates among girls under five years old in Nepal, which suggests the disadvantageous treatment of girls compared with boys at the postnatal stage.\cite{763172:18189218} Second, the technology of sex-selective abortion is accessible to the Nepalis. Even though prenatal sex determination and abortion based on sex selection are prohibited and may incur severe penalties in Nepal under the current abortion law passed in 2002,\cite{763172:18697445,763172:18750509} studies show an emergence of sex-selective abortions being conducted in some areas of Nepal.\cite{763172:18750595} Even before abortion was made legal, Nepali women practiced sex-selective abortion across the border in India.\cite{763172:18752116} Nepal and India share an open border, allowing the free movement of people. The majority of Nepal's population live either near the border with India or within easy access via roads. Third, fertility has been declining in Nepal, with the total fertility rate (TFR), or children per woman, falling from 6.0 in 1950 to 1.9 in 2019 according to the UN.\cite{763172:18699170} The fertility rate in Nepal is expected to decline further since female educational attainment is increasing (e.g. among 20- to 24-year-olds, the proportion of females with at least lower secondary school educational attainment increased from 41\% in 2005 to 57\% in 2015),\cite{763172:18874449} and there is a negative relationship between education and fertility. In 2016, the TFR was 1.8 among women who had completed the tenth grade and 3.3 among those without education.\cite{763172:18874492}

It is crucial to monitor and project the SRB at the subnational level for Nepal, since the national level may mask SRB imbalances in subpopulations. No prior study has found evidence of an ongoing sex ratio transition with imbalanced SRB in Nepal on the national level. However, studies have suggested that sex-selective abortion may exist in certain areas on the subnational level.\cite{763172:18188972,763172:18189027,763172:18189176} On the subnational level, Nepal has high levels of heterogeneity in terms of its demographics, socioeconomics, and culture (i.e. religion, caste, and ethnicity). The majority of Nepalis are Hindus (81\%), followed by Buddhists (9\%), Muslims (4.4\%), Kirats (3\%), Christians (1.4\%), and others.\cite{763172:19214265} Moreover, Hindus are further divided by caste, namely, Brahman, Chhetri, Vaisya, Shudra, and Dalit. Furthermore, more than one religion is practiced within each ethnic group, and the distribution varies. Altogether, there are 125 caste and ethnic groups, with 123 spoken languages; 19 languages are spoken by 96\% of the population.\cite{763172:19214265} In Nepal, some caste and ethnic groups (i.e. Dalit and many indigenous communities called, in Nepali, \textit{Adibasi} and \textit{Janajati}) are defined by the constitution as marginalized populations, with lower socioeconomic status and a history of subjection to oppression compared to higher castes (e.g. Brahman/Chhetri among the Khas ethnicity in the Hills and Madhesi ethnicity in Terai) and ethnicity.\cite{763172:19220071} As a result, not only do fertility levels differ between geographical locations and urban–rural areas,\cite{763172:18874492} the son preference intensity also differs across castes, ethnic groups, religions, and cultures.\cite{nanda2012study} The disadvantaged non-Dalit Terai castes have the highest level of son preference among men, whereas the least son preference is reported among the Brahman/Chhetri.\cite{nanda2012study}

Nepal established a new federal system with seven provinces (called Pradesh) and 753 municipalities in September 2015, with the first elections for the federal and provincial parliaments held in 2017. The seven provinces\footnote{To date, four province-level parliaments have decided the names of their provinces. In this study, we use Province 1 to 7 to label the seven provinces and provide the province names if available.} each act as prime administrative units for policymaking. Most provinces include multiple ecological zones: (i) Mountain in the northern Himalayan belt, (ii) Hill in the central region, and (iii) Terai in the southern plain. Exceptions are Province 2, entirely in Terai; Province 5, which lacks Mountain; and Province 6, \textit{Karnali Pradesh}, which lacks Terai. Table~\ref{tab_popdist} presents Nepal’s provincial profiles, including population distribution by ecological zones, by caste/ethnicity and religion, GDP per capita, and education status. It is crucial to estimate and project the SRB imbalance in Nepal by province to directly assist the monitoring of population indicators and program planning, in order to address the huge heterogeneities across provinces in demography, education, ecology, and culture as shown in Table~\ref{tab_popdist}.

This study aimed to estimate the levels and trends in SRB for the seven Nepal provinces between 1980 and 2016 and provide probabilistic SRB projections up to 2050. The paper is organized as follows: we first explain the data sources and the Bayesian statistical model for SRB estimation and projection. We then present the modeled results for SRB by province over time. In the discussion, we summarize the main contributions, limitations due to data availability and model assumptions, and suggestions for future research.

\section*{Data and methods}
The data and method details, including the preprocessing of data, motivations of model choices, model specifications, priors, statistical computation, and validation approaches and their results are available in the Additional file 1. We summarize the main steps in the rest of the section.

\subsection*{Data }We compiled an SRB database for the seven provinces. The database consists of 91 province-level SRB observations with reference years ranging from 1976 to 2016. The total number of births involved in constructing the SRB database is 151,663. Table~\ref{tab_database} summarizes the database by data source.

We obtained the SRB observations from survey and census microdata\footnote{In this study, we use the term ``microdata'' to refer to individual-level data collected in surveys and censuses.}. For the survey data, we included the 2001, 2006, 2011, and 2016 Nepal Demographic and Health Surveys (NDHSs). ). NDHSs are retrospective surveys which obtain full birth histories (all births that a woman has given in her life so far) from women aged 15--49. We did not include the 1996 NDHS or the 1976 World Fertility Survey because the microdata files do not contain the respondents' district information; we were unable to extract SRB by province from these two surveys. To obtain NDHS observations with associated uncertainty at a controlled level, we merged observations with initially one-year period into a multi-year period based on the coefficient of variation for SRB (Additional file 1 section 1.2).\cite{763172:18220988} We then computed the sampling errors for the observations using a jackknife method for the four NDHSs to reflect the uncertainty resulting from the multi-stage stratifying sample design (Additional file 1 section 1.2).\cite{763172:18252130,763172:18252133,763172:18252378} We excluded NDHS observations that are more than 25 years prior to the survey year to minimize recall bias for the sex composition of older females' full birth history. We included the 2011 Census in the database where the birth file included the sex of the last birth within 12 months prior to the census. The microdata sample from the census consists of around 15\% (4 million individuals) of the population, with a coverage range of 11\%–99\% of the population from each district. When aggregating the district-level birth data to the provincial level, we applied weights to the births data by district according to the proportion of the sampled population by district to adjust for over- and under-sampling. For both the NDHS and the census data, to obtain provincial SRB observations, we aggregated the sex-specific births from 75 districts according to the province they belonged to (Additional file 1 section 1.1), except for the 2016 NDHS, since it already contained province information in the microdata file. We excluded the 2001 Census from our analysis due to its implausible data quality at the district level.

We compiled the TFR (as model input, which will be explained in Methods) by province from the 2001, 2006, 2011, and 2016 NDHSs (Additional file 1 section 1.3) using \textbf{R}-package \texttt{DHS.rates}.\cite{763172:18360179,763172:18360180} For the 2001 NDHS, because the survey respondents were ``ever-married females'' and those who were never married at the time of the survey were not included, we weighted each female respondent using an ``all-women factor'' so that the resulting microdata represented responses from all women instead of only females who were ever married.\cite{763172:18220726} The TFR for the period beyond 2016 was based on a medium-fertility scenario of the population projections for the 753 municipalities.\cite{763172:18875184} We then aggregated the municipality births and the women’s years of exposure to the reproducible periods by province to generate the provincial TFR.

\subsection*{Methods}We used a Bayesian hierarchical time series model to estimate and project the SRB by province, with a focus on modeling the potential SRB imbalance due to sex selection up to 2050. The model is largely based on the one as described in a prior study.\cite{763172:19021860} However, we made a few modifications to the model in this study to produce improved fit for the provincial SRB observations in Nepal. We briefly summarize the model here. The full model details are in Additional file 1 section 2. The sensitivity analyses are in Additional file 1 section 5, where we show that the model results are not sensitive to the model settings.

We define $\Theta_{p,t} $ as the SRB for a Nepal Province $p $ in year $t $. We model $\Theta_{p,t} $ as:
\begin{equation}
\Theta_{p, t} = b \Phi_{p,t} + \delta_p \alpha_{p,t},
\end{equation}
where $b=1.049$, taken from a previous study,\cite{763172:18187947} is the SRB baseline level for the entire country, assumed to be known and constant across provinces over time; $\Phi_{p,t} $ captures the natural fluctuations around the national SRB baseline $b $ and is modeled as a first-order AR(1) on the log scale; $\delta_p $ is the province-specific binary parameter that detects the absence or presence of SRB inflation: $\delta_p=0 $ refers to no SRB inflation (relative to the national baseline $b $) for province $p $; and $\delta_p=1 $ indicates the existence of imbalanced SRB. We assume that $\delta_p $ follows a Bernoulli distribution with province-specific means. The SRB inflation probability by province was computed as the number of posterior samples with $\delta_p=1 $ divided by the total number of posterior samples. $\alpha_{p,t} $ models the SRB inflation process over time for province $p $ in year $t $; it was assumed to be non-negative and to capture the upward skewed SRB due to sex-selective abortion.\cite{763172:18188574} Assuming a trapezoid function according to the study\cite{763172:18187947}, we modeled the period lengths of three stages of a sex ratio transition (i.e. increase, stagnation, and convergence back to the national SRB baseline) and the maximum level of SRB inflation. All the shape parameters related to the trapezoid function (i.e. start year, period lengths of three stages of SRB transition, and maximum level of imbalance) were province-specific and followed hierarchical distributions. We assigned informative priors to the mean and variance of these hierarchical distributions, which were based on the national-level sex ratio transition experience in Nepal.\cite{763172:19021860} The province-specific start year of the SRB inflation incorporated the fertility squeeze effect by using the TFR. The start year parameter followed a Student-\textit{t} distribution with three degrees of freedom. The choice of a heavy-tail distribution is to enable the model to capture the start years of an SRB imbalance with potential outlying TFR among the provinces. The mean of the start year distribution followed the relationship between national SRB imbalance and fertility decline.

The model took into account varying uncertainties associated with the observations. For the observed SRB on the log scale, $\log(r_{p,t}) $, at province $p $ in year $t $, we assumed that it follows a normal distribution:
\begin{equation}
\log(r_{p,t})\vert\Theta_{p,t}\sim\mathcal N(\log(\Theta_{p,t}),\sigma _{p,t}^{2}),
\end{equation}
where $\sigma _{p,t} $ is the sampling error for $\log(r_{p,t}) $, which reflects the uncertainty in the observations due to survey sampling design. It was pre-computed using a jackknife method as explained in the Data section (and in details in Additional file 1 section 1.2).

The model performance was assessed using an out-of-sample validation exercise and a simulation analysis (Additional file 1 section 3). Due to the retrospective nature of the SRB data and the occurrence of data in series, we left out 20\% of the observations. Instead of randomly selecting the left-out observations, we select those data that were collected since 2016 in the out-of-sample validation exercise. This approach has been used in other studies to validate the model's predictive power for demographic indicators based on retrospective information.\cite{763172:18189218,763172:18486706,763172:18486882,alkema2014global} The validation and simulation results suggested that our model was reasonably well-calibrated, with generally conservative credible intervals (Additional file 1 section 4). The sensitivity analyses show that the model results are not sensitive to model assumptions (Additional file 1 section 5).

\section*{Results}
The SRB estimates and projections from 1980 to 2050, including uncertainty for the seven Nepali provinces, are presented in the Additional file 1.

\subsection*{Levels and trends in SRB between 1980 and 2016 by province}Figure~\ref{fig_srb_esti} provides an overview of the estimated levels and trends in SRB for the seven provinces from 1980 to 2016. The modeled estimates imply more disparities in the provincial SRB after 2000 than before 2000. In 2016, the estimated SRB is the highest in Province 5, at 1.102 with a 95\% Bayesian credible interval (1.044, 1.127), and it is the lowest in Province 2, at 1.053 (1.035, 1.109) as shown in Figure~\ref{fig_srb_esti_pt}. None of the provinces has an SRB that is statistically significantly different from the national SRB baseline level of 1.049 (i.e. the baseline SRB for the whole of Nepal was taken from \cite{763172:18187947}). Before 2000, the SRB remains around the national SRB baseline with minor natural fluctuations in all provinces. In 1980, the SRB ranges from 1.047 (1.031, 1.064) in Province 7 (\textit{Sudurpaschhim Pradesh}) to 1.053 (1.036, 1.103) in Province 4 (\textit{Gandaki Pradesh}).

\subsection*{SRB imbalances by province}Table~\ref{tw-8cf8c2b3dca4} summarizes the modeled results of the SRB imbalances for the seven provinces. The start years for the SRB imbalances are estimated to range from 2001 in Province 5 with a 95\% credible interval (1992, 2022) to 2017 (1998, 2040) in Province 2. The TFR at the start of the SRB inflation ranges from 2.6 (i.e. the average number of children born per woman) in Province 2 and 2.7 in Province 6 to as high as 3.9 in Province 7 and 4.4 in Province 5.

The probabilities of having an SRB imbalance varies for all provinces, but it is generally low, ranging from 16\% and 35\% in Province 2 and Province 6 (\textit{Karnali Pradesh}), respectively, to 81\% in Province 5. The average inflation probability for the seven provinces is 53\%. These findings are in line with previous studies that found no SRB imbalance on the national level.

\subsection*{Probabilistic projections of SRB between 2016 and 2050 by province}During the period 2016–2050, the levels, trends, and imbalances of SRB in Nepal are projected to differ across the provinces, given the model assumptions of province-specific probabilities having SRB inflation (see Figure~\ref{fig_srb_proj} and Figure~\ref{fig_srb_proj_map}). At the beginning of the projection period since 2016, the SRB imbalances are projected to start declining to the national SRB baseline in Province 5, 3, and 7. The sex ratio transitions in Province 1 and 4 are in the midst of climbing to the maximum levels of SRB imbalance, and the SRB inflation has just started in Province 2 and 6. The year in which the projected SRB reaches its maximum ranges from 2016 in Province 5, with an SRB of 1.102 (1.044, 1.127), to 2033 in Province 2, with an SRB of 1.074 (1.036, 1.122). Around the mid-2030s, the SRB in all the provinces is projected to start converging back to the SRB national baseline. By 2050, the SRB in all the provinces is projected to be around the baseline level.

\section*{Discussion}
To the best of our knowledge, this is the first study to estimate and project SRB by Nepal province from 1980 to 2050. The database of province-level SRB in Nepal used for this study is by far the most extensive to date; it includes four NDHSs and one census, covering a total of 151,663 birth records. We adopted a Bayesian hierarchical time series model from a prior study (with modifications) to capture natural sex ratio fluctuations around the national baseline for each province, model the sex ratio transition with a province-specific probability of having SRB inflation, and account for varying uncertainties associated with the observations.\cite{763172:19021860} The model captured regularities in sex ratio transition patterns across the provinces and incorporated the TFR to estimate the start year of the sex ratio transition process to capture the fertility squeeze effect. With the Bayesian hierarchical setup, we were able to use information about the national-level sex ratio transition from a previous study to assist in estimating provincial-level SRB imbalances.\cite{763172:19021860} Based on the model assumption of the province-specific probability of having an SRB imbalance, we projected the sex ratio transitions and resulting imbalanced SRB across the provinces. Consequently, the SRB projections for the seven provinces are model-based and data-driven.

The modeled results imply that there have been more disparities in SRB across the provinces since 2000, even though the provincial SRBs are not estimated or projected to be statistically significantly different from the national baseline. We estimate that the probability of having an existing SRB imbalance varies greatly among the seven provinces, from 16\% in Province 2 to 81\% in Province 5. In Province 5, SRB inflation is estimated to start when the TFR declines to 4.4, which is the highest TFR at the start of the sex ratio transition among all the provinces. These findings make Province 5 unique in the context of sex ratio imbalance at birth. A possible explanation for this is its geographical location; the province borders India, a country with a strong son preference and an ongoing SRB imbalance. In addition, Province 5 contains mostly Terai (72\% of the population, as of 2011), and the son preference is relatively higher among Terai ethnic populations compared to the Hills.\cite{nanda2012study} According to a previous study, the TFR was around 5.2 when the SRB in India started to inflate.\cite{763172:18187947} As for Province 2, it has the lowest probability of experiencing an ongoing SRB imbalance according to our results. Although it is geographically and ethnically close to Bihar (100\% of the population reside in Terai, and 83\% of the population are Madhesi, Muslim, or Dalits), an Indian state with a strong son preference and imbalanced SRB,\cite{763172:18188439} Province 2 is one of the least developed provinces in Nepal, given it has the least per capita income (see Table~\ref{tab_popdist}). It could be that abortion technology is not as affordable or accessible in Province 2 as it is in richer provinces. This speculation is in line with studies that have shown that although the established fee for an abortion service in the public sector is around US\$8--14,\cite{763172:19219237,763172:19219238} hidden costs for medications, materials (such as surgical gloves), and equipment (such as syringes), may be beyond the means of poor or marginalized women.\cite{763172:19219450}

Our SRB estimates and probabilistic projections are based on several modeled assumptions and are, therefore, subject to limitations. First, in the sex ratio transition model, we only incorporated the fertility squeeze effect; we did not incorporate any additional factors that may affect SRB imbalance. Several studies have considered the son preference effect on SRB inflation,\cite{763172:18221841,763172:18221844,763172:18221845} but they are either simulation studies that do not estimate and project SRB or the proxy indicator for son preference intensity is based on much bigger population sizes. When looking at these indicators by Nepali province, the values are not informative enough since the sample sizes are too small. Second, instead of modeling global parameters related to the natural fluctuation of SRB in the time series model and global parameters (i.e. not province-specific parameters) related to the sex ratio transition process, we borrowed such information from prior studies.\cite{763172:18187947,763172:19021860} When we attempted to model all these global parameters, the resulting SRB had too much uncertainty to provide any meaningful trends. Hence, we focused the model on the provincial SRB imbalance and assumed the national sex ratio transition experience followed from previous studies. Third, we model that the SRB imbalance is nonnegative because we assume that the SRB imbalance mainly happens in the context of son preference instead of girl preference. Given that Nepal is a country with strong son preference, we do not consider the possibility of SRB deflation. Forth, the uncertainty bounds for the projections in some of the province-years are wide, even after we used informative priors for sex-ratio-transition-related parameters. This was mainly due to the small birth sample sizes for each province from data sources, which resulted in relatively large uncertainties for the observations. Lastly, when interpreting the projected SRB, it is important to bear in mind that the projection was made under the assumption that the SRB will inflate in the future and follow the national experience of sex ratio transition with a probability. Applying other model assumptions may result in slightly different trajectories for projected SRB as we have shown in our sensitivity analyses (Additional file 1 section 5). However, none of these difference is statistically significantly different from zero, for both the estimation and projection periods.

Contrary to earlier centralized policymaking in Nepal, the new federal system has devolved the health sector to provincial- and municipality-level governments. While the system is under development, we can expect greater differences in policies, implementation, and responses from individuals. Our results show that SRB levels and trends and the probabilities of SRB inflation vary greatly across the provinces. Therefore, it is essential to strengthen existing policies and devise new ones, considering the multiple layers of governance in the new federal system.

Future studies could make use of the projected SRB and calculate the number of missing female births to quantify the effect of the imbalanced SRB by Nepali province when the number of births by province are made available.\cite{763172:18343495} As we estimated the probability of SRB inflation for each province, in-depth studies of provinces and municipalities, conditioning on the availability of reasonably good-quality data, are required to monitor and confirm whether a sex ratio imbalance at birth has occurred or is ongoing. Future research, including field studies in collaboration with the government(s) and NGOs/INGOs, would be useful for collecting high-quality data from subpopulations to better monitor prenatal sex discrimination in Nepal.

\section*{List of abbreviations}
\begin{table}[h!]
\begin{tabular}{cc}
SRB & sex ratio at birth\\
TFR & total fertility rate\\
NDHS & Nepal Demographic and Health Survey\\
\end{tabular}
\end{table}

\begin{backmatter}
\section*{Availability of data and materials}
Additional file 1 (technical appendix) is available from the figshare repository, DOI:\href{https://doi.org/10.6084/m9.figshare.12593651}{10.6084/m9.figshare.12593651}. The DHS microdata files are available from the DHS Program website. The datasets supporting the conclusions of this article are included within the article and its additional file. The final database in excel format will be made public once the paper is accepted.

\section*{Competing interests}
  The authors declare that they have no competing interests.

\section*{Funding}
  FC and HO are supported by baseline research grant from King Abdullah University of Science and Technology. SKC is partially supported by the Major Program of the National Social Science Fund of China (Grant No. 16ZDA088).

\section*{Author's contributions}
    FC proposed and conceptualized the study. FC and SKC constructed the SRB database. FC and SKC oversaw the study design. FC developed the statistical model. FC wrote the first draft and the technical appendix. FC, SKC and HO analyzed the results. FC, SKC and HO edited and revised the manuscript.

\section*{Acknowledgements}
  The authors are grateful to Leontine Alkema and Christophe Z. Guilmoto for their valuable comments and discussion on the earlier version of this manuscript. We are thankful for Ryan Rylee's copy editing service.

\bibliographystyle{bmc-mathphys} 
\bibliography{nepal_srb}      




\clearpage
\listoffigures

\section*{Figures}
  \begin{figure}[h!]
  \includegraphics[width = 0.95\linewidth]{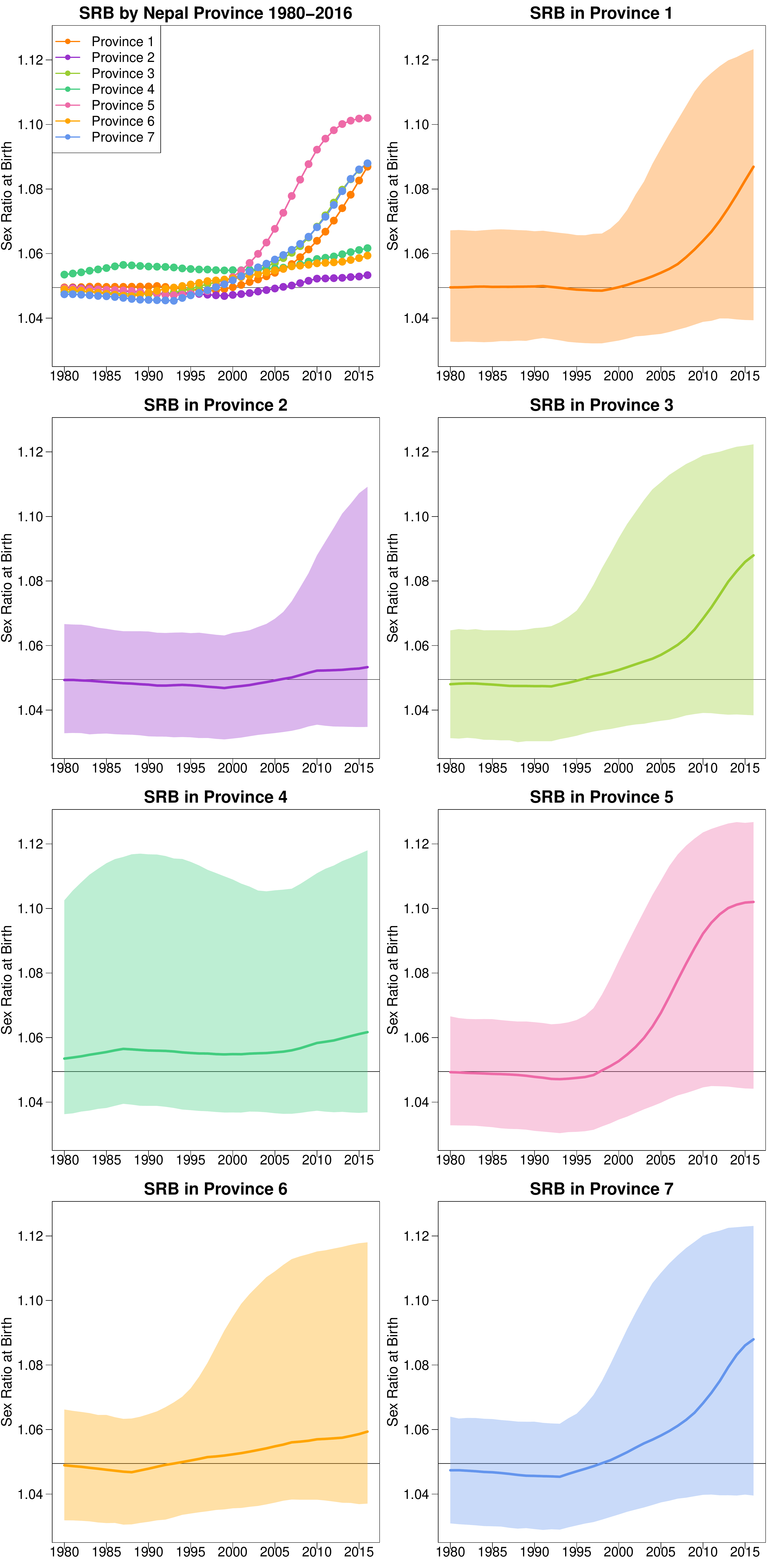}
  \caption[SRB estimates for seven Nepal provinces during 1980--2016]{\csentence{SRB estimates for seven Nepal provinces during 1980--2016.}
      The median estimates are shown in the top left panel across all provinces. In all the rest panels, median estimates are in curves and 95\% credible bounds are in shades. Horizontal line indicates the national SRB baseline value at 1.049.\cite{763172:18187947}}
\label{fig_srb_esti}
      \end{figure}

\begin{figure}[h!]
  \includegraphics[width = 0.95\linewidth]{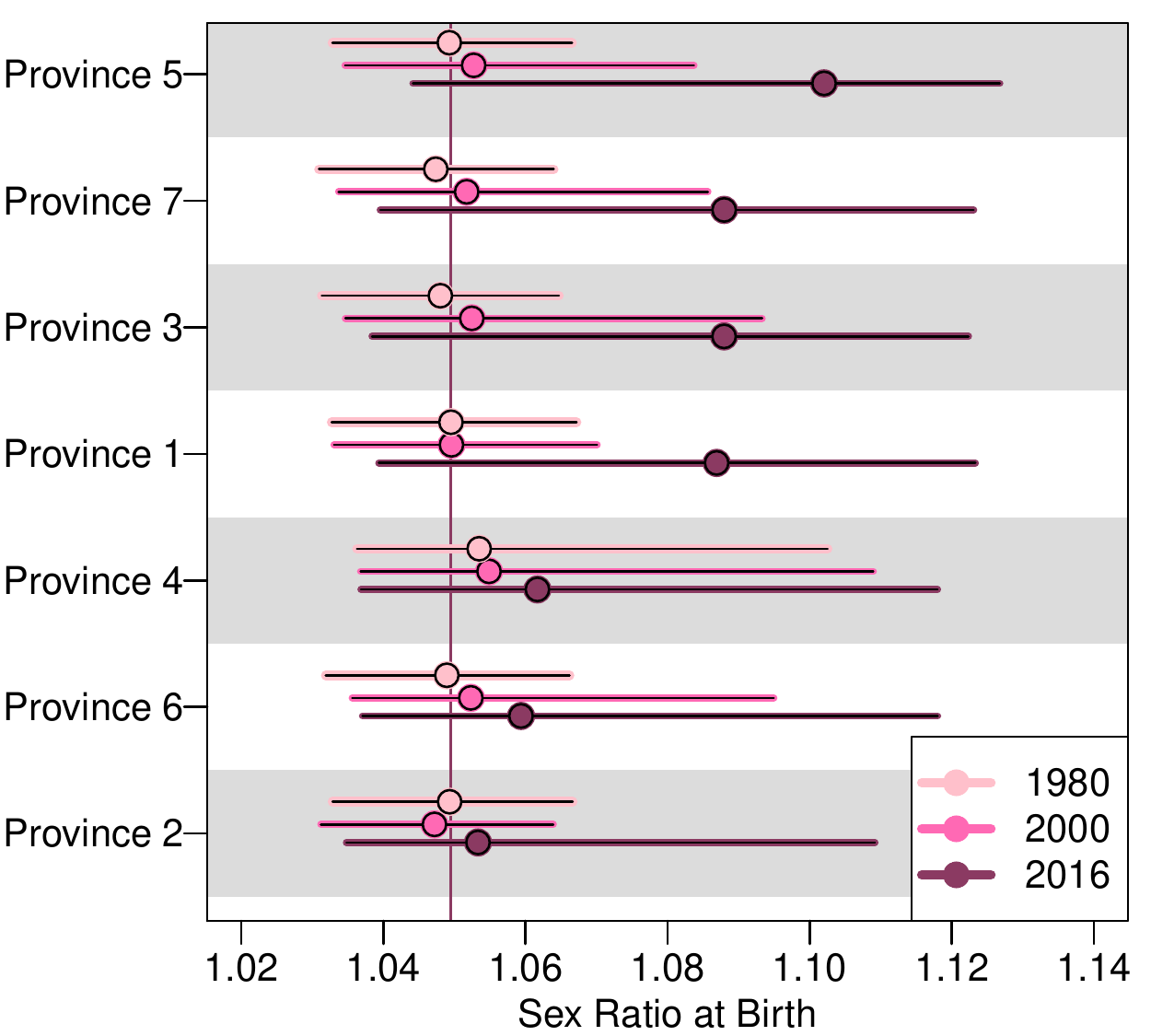}
  \caption[SRB by Nepal province in 1980, 2000 and 2016]{\csentence{SRB by Nepal province in 1980, 2000 and 2016.}
      Dots refer to the median estimates. Line segments are the 95\% credible intervals. Vertical line indicates the national SRB baseline value at 1.049.\cite{763172:18187947} Provinces are in descending order of the median estimates of SRB in 2016.}
\label{fig_srb_esti_pt}
      \end{figure}

\begin{figure}[h!]
  \includegraphics[width = 0.95\linewidth]{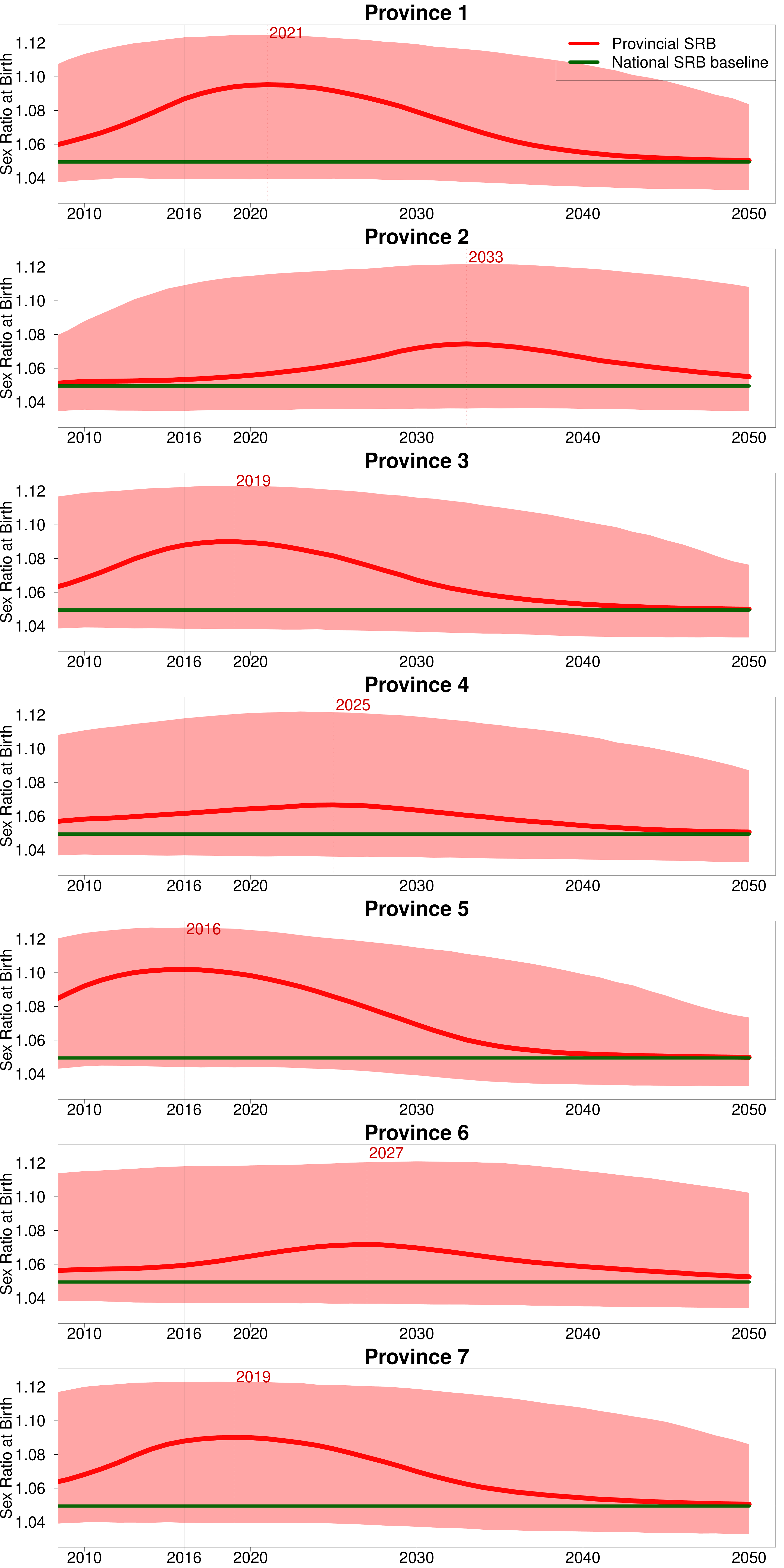}
  \caption[SRB projections for 7 Nepal provinces during 2016--2050]{\csentence{SRB projections for 7 Nepal provinces during 2016--2050.}
      Median projections of provincial SRB (red curve), 95\% credible interval (red shade), national SRB baseline value at 1.049 (green horizontal line).\cite{763172:18187947} The year in which the median projection reaches the maximum is shown.}
\label{fig_srb_proj}
      \end{figure}

\begin{figure}[h!]
  \includegraphics[width = \linewidth]{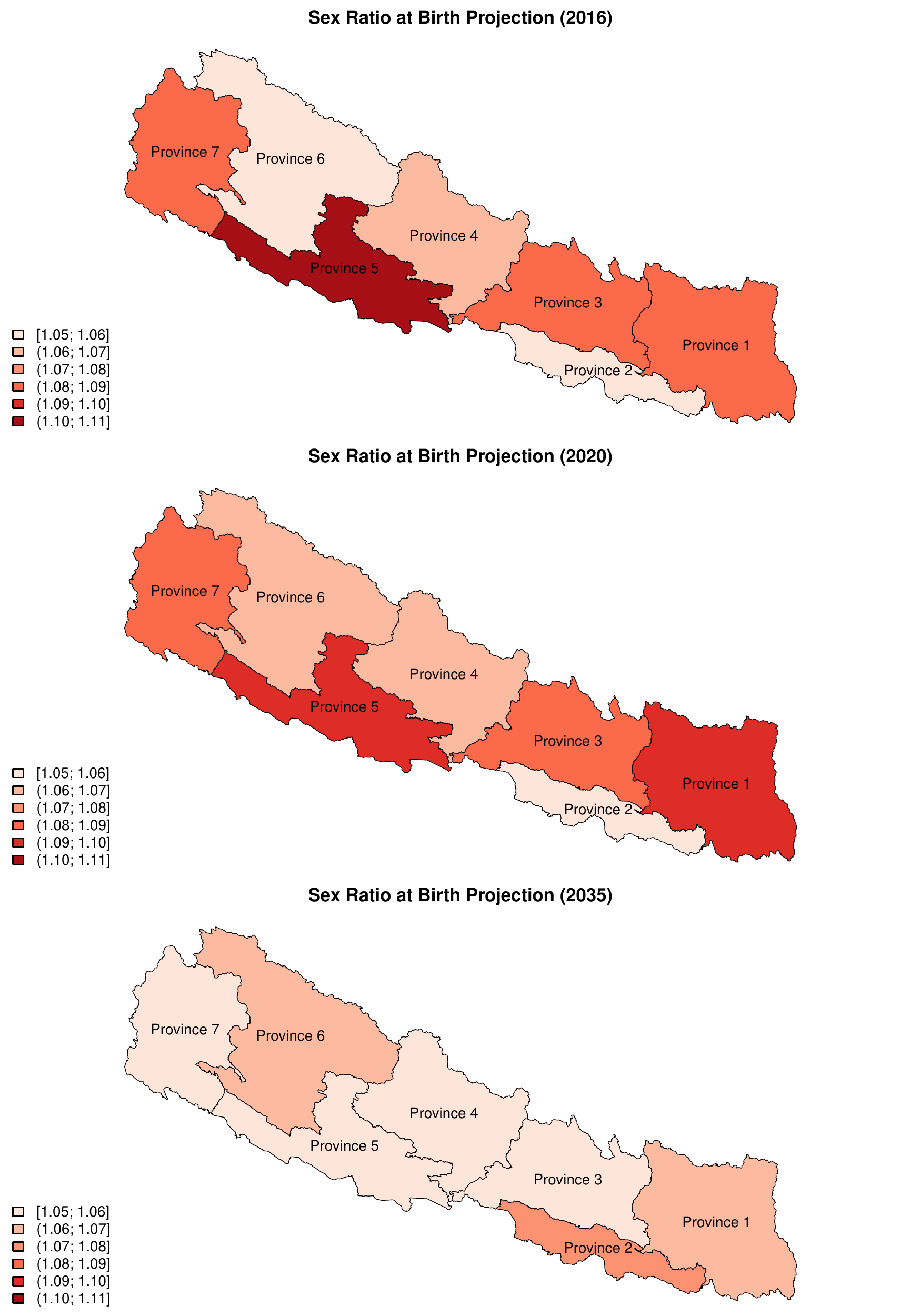}
  \caption[Projected SRB by Nepal province.]{\csentence{Projected SRB by Nepal province.}
      The median SRB projections are shown by Nepal province in 2016 (top), 2020 (middle) and 2035 (bottom).}
\label{fig_srb_proj_map}
      \end{figure}


\clearpage
\section*{Tables}
\begin{table}[h!]
\caption{Demographic, socioeconomic and cultural characteristics by Nepal Province. The population distribution by ecology zone, education attainment, caste/ethnicity and religion are from Census 2011. Educated young adult refers to 20–39 years old with education level at least lower secondary. GDP during 2018–2019 is from \cite{763172:19215390}, using an exchange rate of 1 USD $=$ 121.6 NPR.}
\label{tab_popdist}
\begin{tabular}{ccccccc}
\hline Nepal Province  & \multicolumn{4}{c}{Population distribution by ecological zones}  & Educated  & GDP per capita  \\
(provicne name) & Total (millions) & Terai & Hill & Mountain & young adults &  (USD) 2018--2019  \\
\hline 
Province 1 & 4.53 & 56\% & 35\% & 9\% & 43\% & 942 \\
Province 2 & 5.40 & 100\% & 0\% & 0\% & 19\% & 657 \\
Province 3 & 5.53 & 10\% & 80\% & 9\% & 49\% & 1,744 \\
  (Bagamati) &&&&&&\\
Province 4 & 2.41 & 13\% & 86\% & 1\% & 45\% & 1,018 \\
   (Gandaki)&&&&&&\\
Province 5 & 4.50 & 72\% & 28\% & 0\% & 31\% & 845 \\
Province 6 & 1.56 & 0\% & 75\% & 25\% & 20\% & 768\\
   (Karnali) &&&&&&\\
Province 7 & 2.55 & 48\% & 34\% & 18\% & 22\% & 761 \\
   (Sudurpashchim) &&&&&&\\
\hline 
\end{tabular}
\begin{tabular}{cccccccc}
\hline Nepal Province & \multicolumn{7}{c}{Population distribution by caste/ethnicity and religion}  \\
(provicne name) & Adivasi-Terai & Janajati-Hill  & Khas Hindu & Dalit Hindu & Madhesi Hindu & Muslim & Others \\
 & & &  (non-Dalits) & &  (non-Dalits) & &\\
\hline 
Province 1  & 9\% & 42\% & 28\% & 9\% & 8\% & 3\% & 1\%\\
Province 2 & 5\% & 5\% & 5\% & 15\% & 55\% & 13\% & 2\%\\
Province 3 & 1\% & 62\% & 31\% & 5\% & 1\% & 0\% & 0\%\\
  (Bagamati) &&&&&&&\\
Province 4 & 1\% & 50\% & 32\% & 16\% & 0\% & 1\% & 0\%\\
   (Gandaki)&&&&&&&\\
Province 5 & 14\% & 19\% & 27\% & 15\% & 17\% & 7\% & 1\%\\
Province 6 & 0\% & 17\% & 64\% & 18\% & 0\% & 0\% & 0\%\\
   (Karnali) &&&&&&&\\
Province 7 & 11\% & 3\% & 64\% & 19\% & 2\% & 0\% & 0\%\\
   (Sudurpashchim) &&&&&&&\\
\hline 
\end{tabular}
\end{table}

\begin{table}[h!]
\caption{SRB database by source. The size of birth samples reported in the table refers to the unweighted total number of births within the 25 years prior to when the surveys were conducted.}
\label{tab_database}
      \begin{tabular}{ccc}
        \hline
Data source & \# SRB observations  & Size of birth samples\\ \hline
NDHS 2001 & 21 & 27,266\\
NDHS 2006 & 21 & 25,952\\
NDHS 2011 & 21 & 26,012\\
NDHS 2016 & 21 & 25,592\\
Census 2011 & 7 & 46,841\\ \hline
\textbf{total} & \textbf{91}  & \textbf{151,663}\\ \hline 
\end{tabular}
\end{table}

\begin{table}[h!]
\caption{SRB imbalance by Nepal province. The median estimates of SRB inflation start year are in front of brackets. The 95\% credible intervals for the start year are in brackets. The TFR values in the median estimates of start year are reported. The province names are in brackets if available}
\label{tw-8cf8c2b3dca4}
      \begin{tabular}{cccc}
        \hline
Nepal Province & SRB inflation start year & TFR in start year & Inflation probability\\ \hline
Province 1 & 2006 (1994, 2028) & 3.1 & 62\%\\
Province 2 & 2017  (1998, 2040) & 2.6 & 16\%\\
Province 3  (Bagmati) & 2004  (1989, 2026) & 3.5 & 63\%\\
Province 4  (Gandaki) & 2006  (1976, 2029) & 3.0 & 55\%\\
Province 5 & 2001  (1992, 2022) & 4.4 & 81\%\\
Province 6  (Karnali) & 2013  (1989, 2035) & 2.7 & 35\%\\
Province 7  (Sudurpaschhim) & 2005  (1991, 2028) & 3.9 & 62\%\\
\hline 
      \end{tabular}
\end{table}


\section*{Additional Files}
  \subsection*{Additional file 1 --- Technical appendix}
    Technical appendix includes details on data preprocessing, model specifications and motivations, statistical computing, validation approaches and results, and sensitivity analyses. It is available from the figshare repository, DOI:\href{https://doi.org/10.6084/m9.figshare.12593651}{10.6084/m9.figshare.12593651}.

\end{backmatter}
\end{document}